\documentstyle[preprint,eqsecnum,aps,epsf]{revtex}

\tightenlines
\begin{document}
\preprint{\vbox{
\hbox{INPP-UVA-99-02} 
\hbox{October, 1999} 
\hbox{hep-ph/9910515}
}}
\draft
\def\vp{{\bf p}}
\def\ko{K^0}
\def\kb{\bar{K^0}}
\def\al{\alpha}
\def\ab{\bar{\alpha}}
\def\be{\begin{equation}}
\def\en{\end{equation}}
\def\bea{\begin{eqnarray}}
\def\eea{\end{eqnarray}}
\def\non{\nonumber}
\def\la{\langle}
\def\ra{\rangle}
\def\epp{\epsilon^{\prime}}
\def\vep{\varepsilon}
\def\to{\rightarrow}
\def\up{\uparrow}
\def\dw{\downarrow}
\def\ms{\overline{\rm MS}}
\def\ums{{\mu}_{_{\overline{\rm MS}}}}
\def\u{\mu_{\rm fact}}

\def\pr{{\sl Phys. Rev.}~}
\def\ijmp{{\sl Int. J. Mod. Phys.}~}
\def\jp{{\sl J. Phys.}~}
\def\mpl{{\sl Mod. Phys. Lett.}~}
\def\prp{{\sl Phys. Rep.}~}
\def\prl{{\sl Phys. Rev. Lett.}~}
\def\pl{{\sl Phys. Lett.}~}
\def\np{{\sl Nucl. Phys.}~}
\def\ppnp{{\sl Prog. Part. Nucl. Phys.}~}
\def\zp{{\sl Z. Phys.}~}

\title{Intrinsic Charm in the Nucleon\\}

\author{X. Song}

\address{Institute of Nuclear and Particle Physics\\
Jesse W. Beams Laboratory of Physics\\
Department of Physics, University of Virginia\\
Charlottesville, VA 22901, USA\\}

\maketitle
\begin{abstract}
The quark flavor and spin structure of the nucleon is discussed 
in SU(4) symmetry breaking chiral quark model. The flavor and spin
contents for charm quarks and anti-charm quarks are predicted and 
compared with the results given by other models. The intrinsic 
charm quark contribution to the Ellis-Jaffe sum rule is discussed. 
\end{abstract}

\bigskip
\bigskip
\bigskip


\newpage

\leftline{\bf I. Introduction}

As suggested by several authors long time ago \cite{dg77,stan81,hm83},
there are so called {\it `intrinsic'} heavy quark components in the
proton wave function. The {\it `extrinsic'} heavy quarks are created on a
short time scale in associate with a large transverse momentum reaction and 
their distributions can be derived from QCD bremsstrahlung and pair
production processes, which lead to standard QCD evolution. The intrinsic
heavy quarks, however, exist over a long time scale independent of any 
external probe momentum. They are created from the quantum fluctuations 
associated with the bound state hadron dynamics. Hence the probability of 
finding the intrinsic heavy quark in the hadron is completely determined
by nonperturbative mechanism. Since the chiral quark model provides a
useful tool in studying the quark spin-flavor structure of the
nucleon in nonperturbative way, we will use this model to discuss the 
heavy quark components of the proton. The quark components from the bottom
and top quarks are negligible in the proton at the scale $m_c^2$ or lower,
hence we only discuss the intrinsic charm (IC) quarks in this paper.

Although the SU(3) chiral quark model with symmetry breaking
\cite{song9705} has been quite successful in explaining the quark flavor
and spin contents in the nucleon, the model is unnatural from the
point of view of the standard model. According to the symmetric GIM model 
\cite{gim70}, one should deal with the weak axial current in the framework
of SU(4) symmetry. It implies that the charm quark should play some role 
in determining the spin and flavor structure of the nucleon.
An interesting question in high energy spin physics is whether the
intrinsic charm exists in the proton. If it does, what is the size of 
the IC contribution to the flavor and spin observables of the proton.
There are many publications on this topic (see for instance
\cite{bs91,hk94,bth95,hsv96,vb96,it97,hz97,mt97,blos98,gol98,amt98,pst98}).

In an ICTP internal report \cite{song98}, we have extended the SU(3) 
chiral quark model given in \cite{song9705} to the SU(4) case and 
obtained main results of the spin and flavor contents in the SU(4)
chiral quark model. In this paper, we will give more detail results
on the contributions to the structure of the proton from the intrinsic 
charm and anticharm quarks. The results are compared with those given 
by other approaches. The intrinsic charm contribution to the Ellis-Jaffe
sum rule is also discussed. 
\bigskip

\leftline{\bf II. SU(4) chiral quark model with symmetry breaking}

In the framework of SU(4), there are sixteen pseudoscalar bosons, a
15-plet and a singlet. The effective Lagrangian describing interaction
between quarks and the bosons is
$${\it L}_I=g_{15}{\bar q}\pmatrix{{G}_u^0
& {\pi}^+ & {\sqrt{\epsilon}}K^+ & {\sqrt{\epsilon_c}}{\bar D}^0 \cr 
{\pi}^-& {G}_d^0& {\sqrt{\epsilon}}K^0 &{\sqrt{\epsilon_c}}D^-\cr
{\sqrt{\epsilon}}K^-& {\sqrt{\epsilon}}{\bar K}^0&{G}_s^0 &
{\sqrt{\epsilon_c}}{D}_s^-\cr 
{\sqrt{\epsilon_c}}{D}^0 & {\sqrt{\epsilon_c}}{D}^+
& {\sqrt{\epsilon_c}}{D}_s^+ &{G}_c^0 \cr }q, 
\eqno (1)$$
where ${G}_{u(d)}^0$ and ${G}_{s,c}^0$ are defined as
$${G}_{u(d)}^0=+(-){{\pi^0}\over{\sqrt 2}}+
{\sqrt{\epsilon_{\eta}}}{{\eta^0}\over{\sqrt 6}}+
{\zeta'}{{\eta'^0}\over{\sqrt 3}}-{\sqrt{\epsilon_c}}{{\eta_c}\over 4}
\eqno (2a)$$
$${G}_s^0=-{\sqrt{\epsilon_{\eta}}}{{2\eta^0}\over{\sqrt 6}}+
{\zeta'}{{\eta'^0}\over{4\sqrt 3}}-{\sqrt{\epsilon_c}}{{\eta_c}\over 4}
\eqno (2b)$$
$${G}_c^0=-{\zeta'}{{{\sqrt 3}\eta'^0}\over 4}+
{\sqrt{\epsilon_c}}{{3\eta_c}\over 4}
\eqno (2c)$$
with
$$\pi^0={1\over{\sqrt 2}}(u\bar u-d\bar d);~~~~~
\eta={1\over{\sqrt 6}}(u\bar u+d\bar d-2s\bar s)
\eqno (3a)$$
$$\eta'={1\over{\sqrt 3}}(u\bar u+d\bar d+s\bar s);~~~~~\eta_c=(c\bar c).
\eqno (3b)$$
The breaking effects are explicitly included in (1) and the SU(4) singlet
term has been neglected. Defining $a\equiv |g_{15}|^2$, which denotes the
transition probability of splitting $u(d)\to d(u)+\pi^{+(-)}$, then 
$\epsilon a$ denotes the probability of splitting $u(d)\to s+K^{-(0)}$. 
Similar definitions are used for $\epsilon_{\eta} a$ and $\epsilon_c a$.
If the breaking effects are dominated by mass differences, we expect 
$0~<~\epsilon_{c} a~<~\epsilon_{\eta} a~\leq~\epsilon a~<~a$. We also
have $0~<~\zeta'^2~<<~1$ as shown in \cite{song9705}.

For a valence u-quark with spin-up, the allowed fluctuations are
$$u_{\up,(\dw)}\to d_{\dw,(\up)}+\pi^+,~~
u_{\up}\to s_{\dw}+K^+,~~
u_{\up}\to u_{\dw}+{G}_u^0,~~
u_{\up}\to c_{\dw}+{\bar D}^0,~~
u_{\up}\to u_{\up}.
\eqno (4)$$
Similarly, one can write down the allowed fluctuations for $u_{\dw}$,
$d_{\up}$, $d_{\dw}$, $s_{\up}$, and $s_{\dw}$. Since we are only 
interested in the spin-flavor structure of non-charmed baryons, the
fluctuations from a valence charmed quark are not discussed here.
 
The spin-up and spin-down quark or antiquark contents in the proton,
up to first order of fluctuation, can be written as
$$n_p(q'_{\up,\dw}, {\rm or}\ {\bar q'}_{\up,\dw}) 
=\sum\limits_{q=u,d}\sum\limits_{h=\up,\dw}
n^{(0)}_p(q_h)P_{q_h}(q'_{\up,\dw}, {\rm or}\ {\bar q'}_{\up,\dw}),
\eqno (5)$$
where $P_{q_{\up,\dw}}(q'_{\up,\dw})$ and $P_{q_{\up,\dw}}({\bar
q}'_{\up,\dw})$ are the probabilities of finding a quark $q'_{\up,\dw}$
or an antiquark $\bar q'_{\up,\dw}$ arise from all chiral fluctuations 
of a valence quark $q_{\up,\dw}$. The probabilities
$P_{q_{\up,\dw}}(q'_{\up,\dw})$ and $P_{q_{\up,\dw}}({\bar q}'_{\up,\dw})$ 
can be obtained from the effective Lagrangian (1). In Table I 
only $P_{q_{\up}}(q'_{\up,\dw})$ and $P_{q_{\up}}({\bar q}'_{\up,\dw})$ 
are listed. Those arise from $q_{\dw}$ can be obtained by using the
relations, 
$$P_{q_{\dw}}(q'_{\up,\dw})=P_{q_{\up}}(q'_{\dw,\up}),~~~
P_{q_{\dw}}({\bar q}'_{\up,\dw})=P_{q_{\dw}}({\bar q}'_{\dw,\up})
\eqno (6)$$.
The notations appeared in Table I are defined as 
$$f\equiv{1\over 2}+{{\epsilon_{\eta}}\over 6}+{{\zeta'^2}\over {48}},~~~
f_s\equiv{{2\epsilon_{\eta}}\over 3}+{{\zeta'^2}\over
{48}}+{{\epsilon_c}\over {16}}
\eqno (7a)$$
and 
$$\tilde A\equiv {1\over 2}-{{\sqrt{\epsilon_{\eta}}}\over 6}
-{{\zeta'}\over{12}},
~~~\tilde B\equiv -{{\sqrt{\epsilon_{\eta}}}\over 3}
+{{\zeta'}\over {12}},~~~
\tilde C\equiv {{2\sqrt{\epsilon_{\eta}}}\over 3}+{{\zeta'}\over {12}},~~~
\tilde D\equiv {{\sqrt{\epsilon_c}}\over 4}.
\eqno (7b)$$
The special combinations $\tilde A$, $\tilde B$, $\tilde C$, and $\tilde
D$ stem from the quark and antiquark contents in the neutral bosons
${G}_{u,d,s,c}^0$ appeared in the effective Lagrangian (1) and
defined in (2a)-(2c). The numbers $fa$ and $f_sa$ stand for the
probabilities of the quark splitting 
$u_{\up}(d_{\up})\to u_{\dw}(d_{\dw})+{G}_{u(d)}^0$ and $s_{\up}\to
s_{\dw}+{G}_s^0$ 
respectively.
 
In the limit $\epsilon_c\to 0$ and change $\zeta'$ to $4\zeta'$, the
$f$ and $f_s$ reduce to the corresponding quantities in the SU(3) case.
We also have
$$
{\tilde A}\to {1\over 3}A_{SU(3)},~~~
{\tilde B}\to {1\over 3}B_{SU(3)},~~~
{\tilde C}\to {1\over 3}C_{SU(3)},~~~
{\tilde D}\to 0
\eqno (8)$$
\bigskip

\leftline{\bf II. Quark flavor and spin contents}

We note that the quark helicity flips in the chiral splitting processes 
$q_{\up,(\dw)}\to q_{\dw,(\up)}$+GB, i.e. the first four processes 
in (4), but not for the last one. In the valence approximation, the
SU(3)$\otimes$SU(2) proton wave function gives 
$$n^{(0)}_p(u_{\up})={5\over 3}~,~~~n^{(0)}_p(u_{\dw})={1\over 3}~,~~~
n^{(0)}_p(d_{\up})={1\over 3}~,~~~n^{(0)}_p(d_{\dw})={2\over 3}~.
\eqno (9)$$

Using (5), (9) and the probabilities $P_{q_{\up,\dw}}(q'_{\up,\dw})$ 
and $P_{q_{\up,\dw}}({\bar q}'_{\up,\dw})$ listed in Table I, we obtain
the quark and antiquark flavor contents 
$$u=2+\bar u,~~~d=1+\bar d,~~~s=0+\bar s,~~~c=0+\bar c,~~~
\eqno (10a)$$
where 
$$\bar u=a[1+\tilde A^2+2(1-\tilde A)^2],~~~
\bar d=a[2(1+\tilde A^2)+(1-\tilde A)^2],~~~
\eqno (10b)$$
$$\bar s=3a[\epsilon+\tilde B^2],~~~\bar c=3a[\epsilon_c+\tilde D^2]
\eqno (10c)$$
From (10b), one obtains
$${{\bar u}\over{\bar d}}=1-{{6\tilde A}\over {(3\tilde A-1)^2+8}}
\eqno (11a)$$
$$\bar d-\bar u=2a\tilde A
\eqno (11b)$$
Similarly, one can obtain $2\bar c/(\bar u+\bar d)$, $2\bar c/\sum(q+\bar
q)$ 
and other flavor observables. It is easy to verify that in the limit
$\epsilon_c\to 0$,
all results reduce to those given in the SU(3) case \cite{song9705}.

For quark spin contents, we have
$$\Delta u={4\over 3}[1-a(\epsilon+\epsilon_c+2f)]-a
\eqno (12a)$$
$$\Delta d=-{1\over 3}[1-a(\epsilon+\epsilon_c+2f)]-a
\eqno (12b)$$
$$\Delta s=-a\epsilon 
\eqno (12c)$$
$$\Delta c=-a\epsilon_c 
\eqno (12d)$$
$$\Delta\Sigma\equiv\sum\limits_{q=u,d,s,c}\Delta
q=1-2a(1+\epsilon+\epsilon_c+f)
\eqno (12e)$$
and 
$$\Delta{\bar q}=0,~~~~~~(q=u,d,s,c) 
\eqno (12f)$$
Comparing to the SU(3) case, a new $\epsilon_c$ term has been included
in $\Delta u$, $\Delta d$ and $\Delta\Sigma$, but there is no change for 
$\Delta s$. In SU(4) chiral quark model, the charm quark helicity 
$\Delta c$ is nonzero and definitely negative. The size of the intrinsic
charm (IC) helicity depends on the parameters $\epsilon_c$ and $a$. We
will see below that the range of $\epsilon_c$ is about 0.1$-$0.3. Since
$a\simeq 0.14$, one has 
$$\Delta c\simeq -0.03
\eqno (13a)$$
The ratio of $\Delta c/\bar c$, however, is a constant 
$${{\Delta c}\over{\bar c}}=-{{16}\over {51}}\simeq -0.314
\eqno (13b)$$
which does not depend on any chiral parameters. This is a special
prediction from the chiral quark model.

In the framework of SU(4) parton model, the first moment of the spin
structure function $g_1(x,Q^2)$ in the proton is
$$\int_0^1g_1^p(x,Q^2)dx={1\over 2}[{4\over 9}\Delta u+{1\over 9}\Delta d+
{1\over 9}\Delta s+{4\over 9}\Delta c]
\eqno (14)$$
which can be rewritten as
$$\int_0^1g_1^p(x,Q^2)dx={1\over {12}}[a_3+{{\sqrt 3}\over 3}a_8-
{{\sqrt 6}\over 3}a_{15}+{5\over 3}a_0]
\eqno (15)$$
where the notations
$$a_3=\Delta u-\Delta d,~~~
a_8={1\over {\sqrt 3}}[\Delta u+\Delta d-2\Delta s],~~~
a_{15}={1\over {\sqrt 6}}[\Delta u+\Delta d+\Delta s-3\Delta c]
\eqno (16a)$$
and 
$$a_0=\Delta u+\Delta d+\Delta s+\Delta c
\eqno (16b)$$ 
have been introduced. 
\bigskip

\leftline{\bf IV. Numerical results and discussion.}

To estimate the size of $\Delta c$ and other intrinsic charm
contributions, we use the same parameter set given in
\cite{song9705}, $a=0.145$, $\epsilon_\eta\simeq\epsilon=0.46$,
$\zeta'^2=0.10$. We choose $\epsilon_c$ as a variable, then other 
quark flavor and helicity contents can be expressed as functions of
$\epsilon_c$. We found that 
$$\epsilon_c\simeq 0.1-0.3
\eqno (17)$$
Our model results, data and theoretical predictions from other 
approaches are listed in Table II and Table III respectively, where 
$\epsilon_c=0.20\pm 0.10$ is assumed. One can see that the fit to
the existing data is as good as in the SU(3) case. 

Several remarks are in order:

(1) The chiral quark model predicts an intrinsic charm component of the
nucleon (${2\bar c}/\sum(q+\bar q)$) around $3\%$, which agrees with the
result given in \cite{bs91} and the earlier number given in \cite{dg77}.
But the result given in \cite{hk94} is much smaller (0.5$\%$) than ours. 

(2) The prediction of intrinsic charm polarization $\Delta c=-0.029\pm
0.015$ from the chiral quark model is very close to the result $\Delta
c=-0.020\pm 0.005$ given in the instanton QCD vacuum model \cite{amt98}.
We note that the size of $\Delta c$ given in \cite{pst98} is about two 
order of magnitude smaller than ours. Hence further investigation in 
this matter is needed. 

(3) We plot the ratio $\Delta c/\Delta\Sigma$ as function of $\epsilon_c$ 
in Fig.1. In the range $0.1<\epsilon_c<0.3$, we have
$${{\Delta c}\over {\Delta\Sigma}}=-0.08\pm 0.05
\eqno (18)$$
which agrees well with the prediction given in \cite{blos98} and is
also not inconsistent with the result given in \cite{amt98}.

(4) For the first moment of the spin structure function $g_1^{(p,n)}$, 
we have included the QCD radiative corrections and the results agree 
well with the data.
 
To summarize, we have discussed the intrinsic charm contribution to
the quark flavor and spin observables in the chiral quark model with
symmetry breaking. The results are compatible with other theoretical
predictions.  
\bigskip

\leftline{\bf Acknowledgments}

The author would like to thank S. Brodsky for useful comments and
suggestions. This work was supported in part by the U.S. DOE Grant No.
DE-FG02-96ER-40950, the Institute of Nuclear and Particle Physics, 
University of Virginia, and the Commonwealth of Virginia.

\begin{table}[ht]
\begin{center}
\caption{The probabilities $P_{q_{\up}}(q'_{\up,\dw},\bar q'_{\up,\dw})$
and $P_{q_{\up}}(q'_{\up,\dw},{\bar q}'_{\up,\dw})$} 
\begin{tabular}{cccc} 
$q'$ &$P_{u_{\up}}(q'_{\up,\dw})$ & $P_{d_{\up}}(q'_{\up,\dw})$ &
$P_{s_{\up}}(q'_{\up,\dw})$ \\ 
\hline 
$u_{\up}$ & $1-({{1+\epsilon+\epsilon_c}\over 2}+f)a+
{a\over {2}}(1-\tilde A)^2$ & ${a\over {2}}\tilde A^2$ 
&${a\over {2}}\tilde B^2$\\
$u_{\dw}$ & $({{1+\epsilon+\epsilon_c}\over 2}+f)a+
{a\over {2}}(1-\tilde A)^2$ & $a+{a\over {2}}\tilde A^2$ &$\epsilon
a+{a\over {2}}\tilde B^2$ \\
$d_{\up}$ & ${a\over {2}}\tilde A^2$ &$1-({{1+\epsilon+\epsilon_c}\over 2}+f)a+
 {a\over {2}}(1-\tilde A)^2$ &${a\over {2}}\tilde B^2$ \\
$d_{\dw}$ & $a+{a\over {2}}\tilde A^2$ &
$({{1+\epsilon+\epsilon_c}\over 2}+f)a+{a\over {2}}(1-\tilde A)^2$ & 
$\epsilon a+{a\over {2}}\tilde B^2$ \\
$s_{\up}$ & ${a\over {2}}\tilde B^2$ &${a\over {2}}\tilde B^2$ &
$1-(\epsilon+f_s+{{\epsilon_c}\over 2})a+{a\over {2}}\tilde C^2$ \\
$s_{\dw}$ & $\epsilon a+{a\over {2}}\tilde B^2$ & $\epsilon a
+{a\over {2}}\tilde B^2$ 
&$(\epsilon+f_s+{{\epsilon_c}\over 2})a+{a\over {2}}\tilde C^2$ \\
$c_{\up}$ & ${a\over {2}}\tilde D^2$ &${a\over {2}}\tilde D^2$ &
${a\over {2}}\tilde D^2$ \\
$c_{\dw}$ & $\epsilon_c a+{a\over {2}}\tilde D^2$ & $\epsilon_c a
+{a\over {2}}\tilde D^2$ 
&$\epsilon_c a+{a\over {2}}\tilde D^2$ \\
\hline
${\bar u}_{\up,\dw}$ &${a\over {2}}(1-\tilde A)^2$ & ${a\over 2}(1+
\tilde A^2)$ &${a\over 2}(\epsilon+\tilde B^2)$ \\
${\bar d}_{\up,\dw}$ &${a\over 2}(1+\tilde A^2)$&
${a\over {2}}(1-\tilde A)^2$ &${a\over 2}(\epsilon+\tilde B^2)$ \\
${\bar s}_{\up,\dw}$ &${a\over 2}(\epsilon+\tilde B^2)$&
${a\over 2}(\epsilon+\tilde B^2)$& ${a\over {18}}\tilde C^2$ \\
${\bar c}_{\up,\dw}$ &${a\over 2}(\epsilon_c+\tilde D^2)$&
${a\over 2}(\epsilon_c+\tilde D^2)$&${a\over 2}(\epsilon_c+\tilde D^2)$
\\
\end{tabular}
\end{center}
\end{table}

\begin{table}[ht]
\begin{center}
\caption{Quark flavor observables}
\bigskip
\begin{tabular}{|c|c|c|c|} \hline
Quantity & Data   & SU(3) &  SU(4)\\
\hline 
$\bar d-\bar u$ & $0.147\pm 0.039$ & 0.147 & 0.120  \\
                & $0.110\pm 0.018$ &       &        \\
\hline 
${{\bar u}/{\bar d}}$ &$[{{\bar u(x)}\over {\bar d(x)}}]_{0.1<x<0.2}
=0.67\pm 0.06$ & 0.65 & 0.69\\
&$[{{\bar u(x)}\over {\bar d(x)}}]_{x=0.18}=0.51\pm 0.06$ & & \\ 
\hline
${{2\bar s}/{(\bar u+\bar d)}}$ & ${{<2x\bar s(x)>}\over {<x(\bar
u(x)+\bar d(x))>}}=0.477\pm 0.051$& 0.69 & 0.69\\
${{2\bar c}/{(\bar u+\bar d)}}$ & $-$ & 0 & $0.28\pm 0.14$ \\
\hline
${{2\bar s}/{(u+d)}}$ & ${{<2x\bar s(x)>}\over
{<x(u(x)+d(x))>}}=0.099\pm 0.009$& 0.128 &0.120\\
${{2\bar c}/{(u+d)}}$ & $-$ & 0 &$0.05\pm 0.02$\\
\hline
$f_s\equiv{2\bar s}/\sum(q+\bar q)$ & $0.10\pm 0.06$ & 0.10&0.09\\
       & $0.15\pm 0.03$ &     &   \\
       & ${{<2x\bar s(x)>}\over {\sum<x(q(x)+\bar q(x))>}}
=0.076\pm 0.022$ & &      \\
$f_c\equiv{2\bar c}/\sum(q+\bar q)$ & 0.03~\cite{bs91} & 0 &$0.03\pm
0.01$\\
& 0.02~\cite{dg77} &  & \\
& 0.005~\cite{hk94}&  & \\
\hline
${{\sum\bar q}/{\sum q}}$ & ${{\sum<x\bar
q(x)>}\over {\sum<xq(x)>}}=0.245\pm 0.005$ & 0.235 & 0.246\\
 $f_3/f_8$ & $0.23\pm 0.05$  & 0.21 &0.22\\
\hline
\end{tabular}
\bigskip
\end{center}
\end{table}

\begin{table}[ht]
\begin{center}
\caption{Quark spin observables}
\bigskip
\begin{tabular}{|c|c|c|c|}\hline
Quantity & Data   & SU(3) &  SU(4)\\
\hline 
$\Delta u$ & $0.85\pm 0.04$ & 0.86 &0.83\\
$\Delta d$&$-0.41\pm$0.04 &$-$0.40&$-$0.39\\
$\Delta s$&$-0.07\pm$0.04 &$-$0.07&$-$0.07\\
$\Delta c$ & $-0.020\pm 0.004$~\cite{amt98} & 0& $-0.029\pm 0.015$ \\
           & $-5\cdot 10^{-4}$~\cite{pst98} &  &          \\
\hline
$\Delta\bar u$, $\Delta\bar d$ & $-0.02\pm 0.11$&0&0\\
$\Delta\bar s$, $\Delta\bar c$ & $-$&0&0\\
\hline
$\Delta c/\Delta\Sigma$ & $-0.08\pm 0.01$~\cite{blos98} & 0& $-0.08\pm
0.05$ \\
  & $-0.033$ ~\cite{amt98}& &  \\
$\Delta c~/~c$ & $-$ &  $-$& $-$0.314  \\
\hline 
$\Gamma_1^p$ & $0.136\pm 0.016$ & 0.133 &0.133\\
$\Gamma_1^n$ & $-0.036\pm 0.007$ & $-0.037$ & $-0.034$  \\
\hline 
$\Delta_3$ &1.2573$\pm$0.0028 &1.26&1.259\\
$\Delta_8$& 0.579$\pm$ 0.025& 0.60&0.578 \\
\hline
\end{tabular}
\bigskip
\end{center}
\end{table}

\begin{figure}[h]
\epsfxsize=5.0in
\centerline{\epsfbox{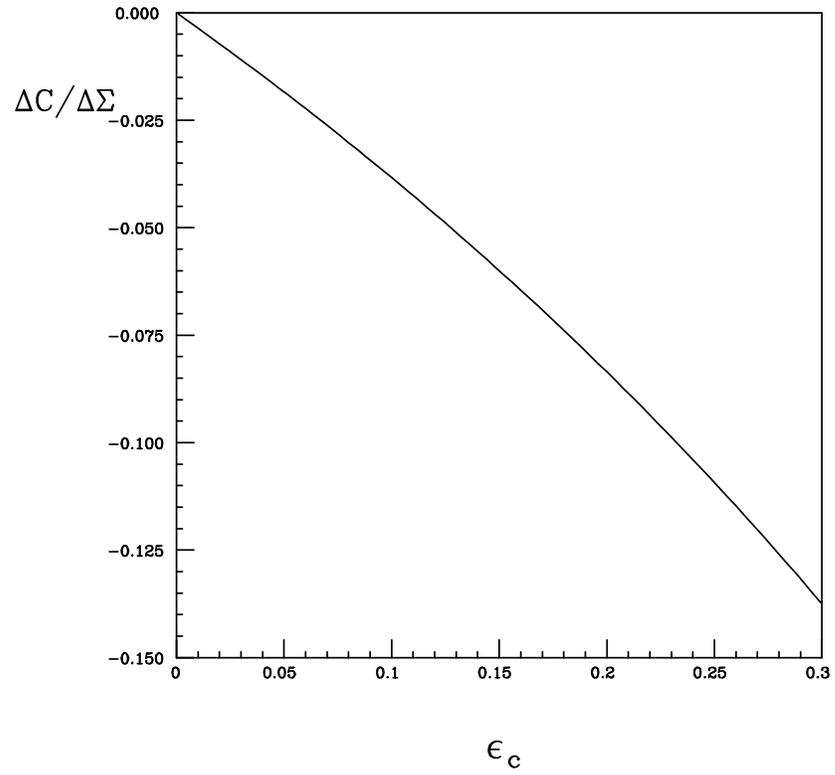}}
\caption{Intrinsic charm quark polarization in the proton as function of
$\epsilon_c$.}
\end{figure}

\end{document}